\documentclass[a4paper]{article}

\usepackage{atmohead2013}
\usepackage[english]{babel}

\title{Determining Atmospheric Aerosol Content With An Infra-red Radiometer}

\shorttitle{Aerosol Content From Sky Temperature}

\authors{
M~K~Daniel$^{1,2}$,
Y~T~E~Lo$^{1}$,
P~M~Chadwick$^{1}$
}

\afiliations{
$^1$ Dept of Physics, University of Durham, Durham, DH1 3LE. UK.\\
\scriptsize{
$^{2}$ now at: Dept. of Physics, University of Liverpool, Liverpool, L69 7ZE. UK.\\
}
}

\email{Michael.Daniel@liverpool.ac.uk}

\abstract{The attenuation of atmospheric Cherenkov photons is dominated by two 
          processes: Rayleigh scattering from the molecular component and Mie 
          scattering from the aerosol component. Aerosols are expected to 
          contribute up to 30\,Wm$^{-2}$ to the emission profile of the 
          atmosphere, equivalent to a difference of $\sim 20^\circ$C to the 
	  clear sky
          brightness temperature under normal conditions. Here we investigate 
          the aerosol contribution of the measured sky brightness temperature 
          at the H.E.S.S. site; compare it to effective changes in the 
          telescope trigger rates; and discuss how it can be used to provide an 
          assessment of sky clarity that is unambiguously free of telescope 
          systematics.}


\begin{document}
\maketitle

\section{Introduction}
The atmosphere is the most important part of the detector in ground-based
gamma-ray astronomy, but it is also the part that has the greatest systematic
uncertainty and over which we have the least control. It falls upon us to
instead monitor and characterise the atmospheric conditions at the time of
observations so that we can either feed this information into Monte Carlo
simulations or reject data when conditions go out of acceptable parameters.

Cherenkov light is generated in the upper atmosphere and will only reach
a ground based Cherenkov telescope if not attenuated through the process of 
Rayleigh scattering on the molecular component of the atmosphere, or Mie 
scattering on the aerosol component (variously dust, silicates, pollens, smoke, 
etc). 
The molecular component of the atmosphere tends to change relatively slowly, 
through seasonal variations; whereas the aerosol component can change much more 
rapidly, depending on wind conditions for example. It becomes vitally important 
to characterise this aerosol component of the atmosphere through regular 
monitoring taken concurrently with telescope observations. A lidar is generally 
used to measure the atmospheric transmission (eg \cite{bib:nolan}) from 
backscattered laser light. At the H.E.S.S. site a lidar centred at 355 and 
532\,nm has been running in conjunction with observations since mid-2011
\cite{bib:vasileiadis}. 
Whilst lidars are excellent instruments for determining the presence of 
aerosols they are not without complications. Firstly a lidar, due to geometric 
viewing considerations, only becomes effective above a minimum altitude, whereas
the aerosol concentration will always be greatest close to the ground.  
Secondly, in order to obtain a transmission profile relevant to the Cherenkov 
spectrum the laser wavelengths are close to the peak in the emission, this 
means the H.E.S.S. lidar is operated only between observing runs to avoid any 
light contamination to the telescope images. In this paper we look at utilising 
another piece of the H.E.S.S. atmospheric monitoring equipment, the all sky
scanning infrared radiometer, to fill in some of this missing information.

The radiometer measures an equivalent brightness temperature for the atmosphere
and correlates this with the atmospheric clarity conditions. 
The atmosphere is split into regions according to its temperature behaviour.
The troposphere is the lowest, most dense, part of the atmosphere -- where
most of the weather happens -- and is characterised by a linear decline in
temperature with increasing altitude and vertical mixing. The molecular density
profile falls off exponentially, with a scale height of a $\sim$few km; the 
vertical air motion in this region also mixes in the larger aerosols which have 
a smaller scale height of order a $\sim$km. The molecular component is an 
inefficient black-body radiator in the 8-14$\mu$m region of the spectrum, 
water vapour and aerosols are slightly more efficient and clouds are very 
efficient. This makes an infra-red radiometer an effective cloud monitor (and
this is the primary function of this instrument), with clouds showing up as
a large brightness temperature compared to a relatively ``cold" sky
\cite{bib:buckley}. H.E.S.S. employ Heitronics KT19.82 radiometers with a
Germanium lens of 2$^\circ$
field of view to monitor for the presence of clouds, with each telescope having
a paraxially mounted unit and a further independent unit continuosly scanning 
the whole sky.

The infra-red luminosity of the sky ($L_{\mathrm{sky}}$) is actually a 
collective sum of the emission of a number of different constituent parts
\begin{eqnarray*}
L_{\mathrm{sky}} = \epsilon_{l}\sigma T_{\mathrm{lens}}^4
                 + \epsilon_{wv}\sigma \int T_{wv}^4
                 + \epsilon_{a}\sigma \int T_{a}^4
                 + \epsilon_{m}\sigma \int T_{m}^4
                 + \ldots
\end{eqnarray*}
where $\epsilon$ denotes the emissivity of the lens ($l$), the water
vapour ($wv$), the aerosols $a$, and the molecular ($m$) profiles of the
atmosphere, respectively and $\int T$ is the relevant integrated temperature profile in 
the line of sight. As the lower altitudes are also the warmest these will
contribute the most to the observed infrared radiation budget, this is also the
region that will be most affected by the varying aerosol concentration. 
According to \cite{bib:dalrymple} the aerosol component can contribute up to 
30\,Wm$^{-2}$ to the bolometric luminosity of a clear sky\footnote{which can 
mean the difference between a brightness temperature of -56$^\circ$C or 
-70$^\circ$C in the presence or absence of aerosols respectively}. This leads 
to the prospect of changing aerosol conditions leading to a noticeable change 
in the sky brightness temperature ($L_{\mathrm{sky}} = \sigma
T^{4}_{\mathrm{sky}}$) measurements.

 \begin{figure*}[!ht]
  \centering
  \includegraphics[width=0.4\textwidth]{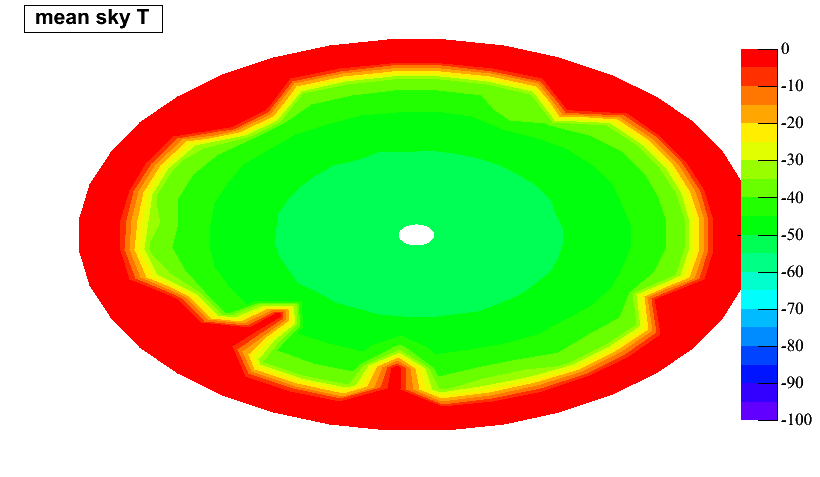}
   \includegraphics[width=0.4\textwidth]{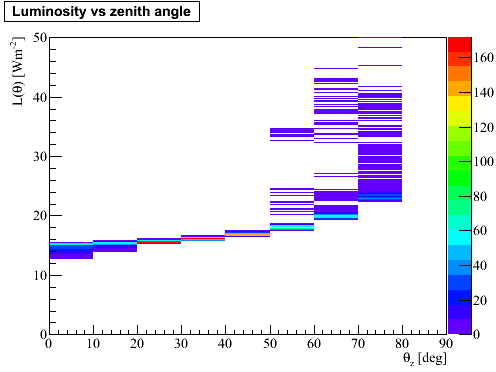}
 \includegraphics[width=0.4\textwidth]{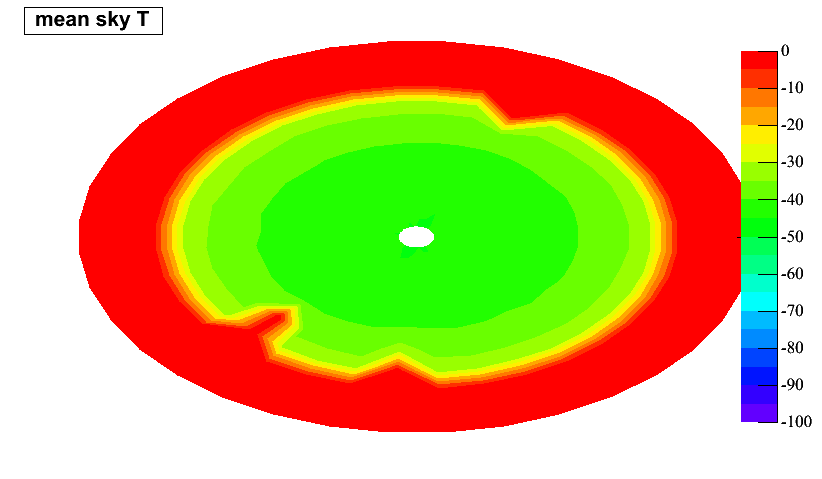}
   \includegraphics[width=0.4\textwidth]{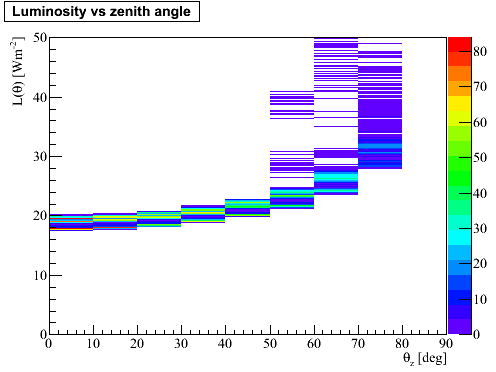}
  \caption{Scanning radiometer sky brightness temperature measurements as a
           function of zenith angle. The top plots are for a clear night with 
	   no discernable aerosol boundary layer; the bottom for a night with 
	   a boundary layer up to $\sim3.5$\,km. The left plots show the all sky
	   temperature values as a function of zenith and azimuth angle, the 
	   right plots are histograms of the accumulated luminosity values as a
	   function of zenith angle.
  }
  \label{fig:T}
 \end{figure*}

 \begin{figure*}[!t]
  \centering
  \includegraphics[width=\textwidth]{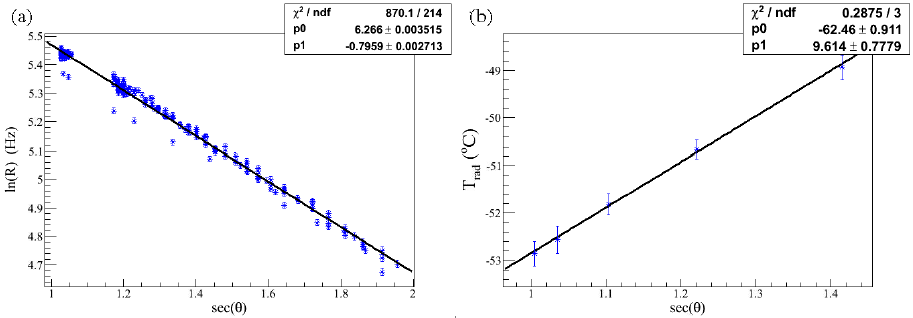}
  \caption{The telescope trigger rate and sky brightness temperature as a
  function of zenith angle for a ``clear sky''.}
  \label{fig:clear}
 \end{figure*}

\section{Data and Observations}
The infrared radiometer readings were compared to the observed telescope 
trigger rates for each night of observations in the period July 2011 to 
July 2012.
\begin{itemize}
\item all four telescopes are functional; 
\item there are no obvious clouds; 
\item there are at least 7 runs with $>35^\circ$ spread in zenith angles in a
night; 
\item full 28 minute observing runs, so that there is no hardware or rate
variations due to sun/moon rise contamination.
\end{itemize}
The zenith angle considerations are needed because the telescopes will not
always point at the same angles each night, the airmass difference at
different observed zenith angles will change both the telescope trigger rate
(due to increased threshold at larger zenith angle) and the brightness
temperature in an inversely correlated fashion. A fit to the rates at a range of
zenith angles allows the rate at a fixed angle (here zenith) to be calculated
and compared to the sky temperature at that fixed angle.
The `ideal' observing conditions at zenith are thus extrapolated from the 
observed measurements. Figure~\ref{fig:T} shows the all sky brightness 
temperature plots taken a few nights apart for two different aerosol conditions, 
the top plot when no aerosol boundary layer is present $>1$\,km in the lidar 
observations and the bottom plot for when there is one to 3.5\,km, a noticeable
shift in the sky's infra-red luminosity can be seen.

To facilitate the comparison, the aerosol conditions were determined from the 
H.E.S.S. lidar runs to split the data into two sets: ``clear sky'' for when 
there is no obvious aerosol boundary layer; and ``hazy sky''.for when the lidar 
observes a boundary layer (up to 3.5\,km under the worst observing conditions).
This leads to 19 ``clear sky'' nights ($\sim100$ runs) with which to empirically
determine the ``clear sky'' telescope trigger rate and sky temperature at zenith 
(eg \ref{fig:clear}) and 28 observing runs under less than ideal conditions.
There are fewer ``hazy sky'' observations from a combination of the site being
astronomically very good and because the telescopes tend not to be operated 
under less than ideal conditions. The correction for the lens temperature and
other systematic effects is taken from measurements of the ambient temperature 
taken by a Campbell Scientific weather station co-located with the scanning 
radiometer.

\begin{figure}[t]
  \centering
  \includegraphics[width=0.4\textwidth]{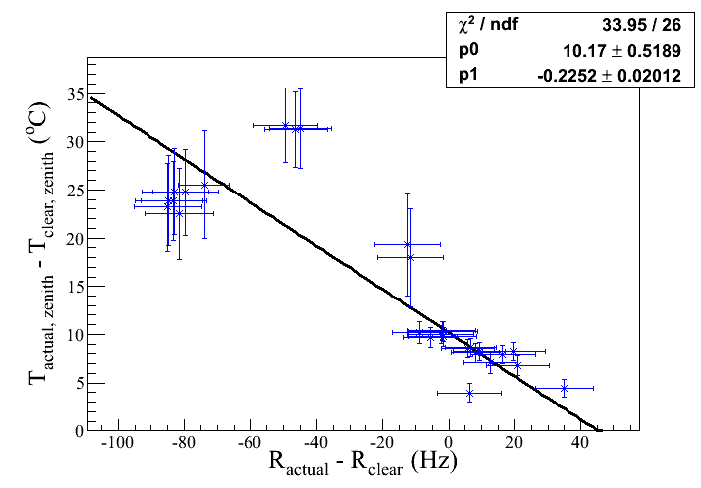}
  \includegraphics[width=0.4\textwidth]{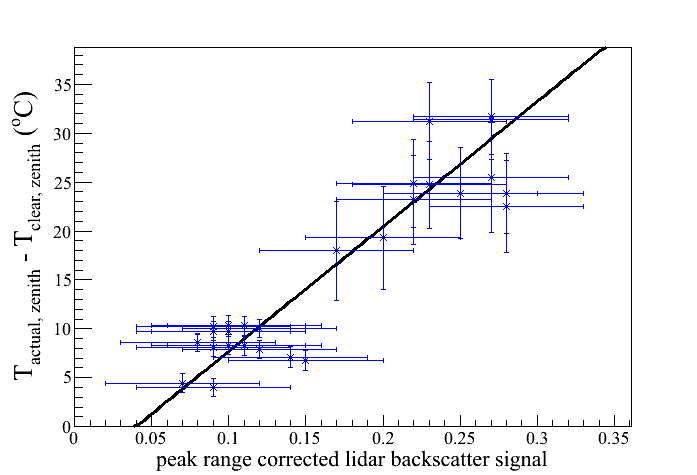}
  \caption{The top figure shows the inverse correlation of the telescope 
  trigger rate with the difference of the observed sky brightness temperature 
  to that expected for a clear sky. The bottom figure shows the correlation
  of the peak in the lidar return signal to the temperature difference.}
  \label{fig:correlation}
 \end{figure}

In figure~\ref{fig:correlation} the difference the measured sky brightness
temperature for the hazy sky runs to that expected from the clear sky runs is
plotted against the difference in the measured and expected telescope trigger 
rates (top plot) and the lidar determined aerosol concentration (bottom plot).
The inverse correlation with telescope trigger rate shows that when the sky has
a higher brightness temperature there is a change in the threshold of the
telescope, as seen by the change in trigger rate 
(Pearson's correlation coefficient of -0.86); the additional positive 
correlation with the aerosol concentration gives us a high level of confidence 
to say the temperature increase is associated with a corresponding increase in 
the atmospheric aerosol content.

\section{Conclusions and Discussion}
The correlation between sky brightness temperature and lidar determined aerosol
concentration has been shown before \cite{bib:daniel} on a limited data set.
Here the study has been expanded to larger, year-long, dataset and once again a
positive correlation of sky brightness temperature and aerosol concentration is
seen. This also corresponds to a clear inverse correlation with Cherenkov
telescope trigger rates. This shows that a radiometer can be a useful instrument 
not just as a cloud monitor, but also as a sensitive measure of atmospheric 
clarity. As the radiometer will be most sensitive to the warmest aerosols it is 
a useful instrument in the low altitude region where a lidar is not sensitive 
due to it not having reached full geometric overlap. An infrared radiometer is 
also a completely passive instrument, allowing it to take fully contemporaneous 
data with a Cherenkov telescope without affecting observations. It can also take 
data during the daytime, allowing a forecast of that night's observing conditions 
to be easily made.

\vspace*{0.5cm}
\footnotesize{{\bf Acknowledgment: }{We would like to acknowledge the support of
the technical staff at Durham and the H.E.S.S. telescopes in the operation of
the radiometer and the people at LUPM in the running of the lidar.}}

\end{document}